\documentclass[namedreferences,article ]{SolarPhysics}
\usepackage[optionalrh,solaenum]{spr-sola-addons}
\usepackage{latexsym}
\usepackage{epsfig}
\usepackage{graphicx}
\usepackage{subfig}
\usepackage{color}           
\usepackage{url}             

\begin{document}

\begin{article}

\begin{opening}
\title{\center{Damping of longitudinal magneto-acoustic oscillations in slowly varying coronal plasma}}

\author{R.~\surname{Erd\'{e}lyi}, K.S.~\surname{Al-Ghafri}
and R.J.~\surname{Morton}}

\runningauthor{R. Erd\'{e}lyi, K.S. Al-Ghafri and R.J. Morton}
\runningtitle{Damping of longitudinal magneto-acoustic oscillations in slowly varying coronal plasma}

\institute{Solar Physics and Space Plasma Research Centre (SP$^2$RC), University of Sheffield,\\ Hicks Building, Hounsfield Road, Sheffield S3 7RH, UK \\ email: \url{[robertus; app08ksa; app07rjm]@sheffield.ac.uk}}
\begin{abstract}
\medskip

We investigate the propagation of MHD waves in a homogenous, magnetized plasma in a weakly stratified atmosphere, representing hot coronal loops. In most of earlier studies a time-independent equilibrium is considered. Here we abandon this restriction and allow the equilibrium to develop as function of time. In particular, the background plasma is assumed to be cooling due to thermal conduction. The cooling is assumed to be on a time scale greater than the characteristic travel times of the perturbations. We investigate the influence of cooling of the background plasma on the properties of magneto-acoustic waves. The MHD equations are reduced to a 1-D system modelling magneto-acoustic modes progressing along a dynamically cooling coronal loop. A time dependent dispersion relation which describes the propagation of the magneto-acoustic waves is derived by using the WKB theory. An analytic solution for the time-dependent amplitude of waves is obtained and the method of characteristics is used to find an approximate analytical solution. Numerical calculations are applied to the analytically derived solutions to obtain further insight into the behavior of the MHD waves in a system with variable, time-dependent background. The results show that there is a strong damping of MHD waves that can be linked to the widely observed damping of hot coronal loop oscillations. The damping also appears to be independent of position along the loop. Studies of MHD wave behaviour in time-dependent background seem to be a fundamental and very important next step in developing MHD wave theory applicable to a wide range in solar physics.
\end{abstract}
\keywords{magnetohydrodynamics(MHD) - plasmas - Sun: corona - waves}
\end{opening}

\section{Introduction}
The solar atmosphere is known to be strongly magnetised and dynamic in nature \cite{Vaiana73,Schrijver99}. The fine structure of this structured plasma environment consists of a wide range of distinct magnetic features like coronal loops, open flux tubes, prominences, etc. These structures play an important role in solar dynamics since they can support various type of MHD waves and oscillations which are thought to contribute to the solution of the long-standing problem of solar coronal heating (for some recent reviews see, \opencite{Klimchuk06}; \opencite{Erdelyi08a}; \opencite{Taroyan08}; \opencite{Taroyan09}). Observations indicate that waves and oscillations are ubiquitous in the solar atmosphere (\opencite{Wang03}; \opencite{Tomczyk07}; \opencite{Pontieu07b}; \opencite{Okamoto07}; \opencite{ErdelyiTaroyan08}; \opencite{Van08}; \opencite{Jess09}. For recent reviews see, \opencite{Nakariakov05}; \opencite{Banerjee07}; \opencite{Moortel09}; \opencite{Zaqarashvili09}; \opencite{Mathioudakis11}). Many of the above reputs highlight that the waves or oscillations detected are seen to be strongly damped. A common suggested mechanisms for damping are resonant absorption \cite{Ruderman,Goossens02,Aschwanden03} and thermal conduction (\opencite{Ofman02}; \opencite{Moortel03}, \citeyear{Moortel04a}; \opencite{Moortel04b}; \opencite{Mendoza04}; \opencite{Erdelyi08}) which cause the damping of fast kink and (slow or acoustic) longitudinal waves, respectively. More recently, Morton and Erd\'{e}lyi (\citeyear{Morton09b}, \citeyear{Morton09c}) argue that radiation of the background plasma may be a dominant damping mechanism for coronal oscillations.

In recent years, coronal waves have been used as a diagonstic tool to infer otherwise unmeasurable or hard to measure plasma parameters e.g. magnetic field \cite{Nakariakov01,ErdelyiTaroyan08} and scale height \cite{Verth08}. Such a technique is known as magneto-seismology and was suggested by \inlinecite{Uchida70}; \inlinecite{Zaitsev83} and \inlinecite{Roberts84} in the context of coronal diagnostics. The concept was proposed to be used to any magnetic structure of the Sun by \inlinecite{Erdelyi06} and labelled as solar magneto-seismology. The models used for deriving the plasma parameters are all static in these earlier works, i.e. the background (or equilibrium) plasma is not dynamic, and background motion with cooling/heating is more than often neglected. However, there are numerous observations which show that the corona, and in particular coronal loops, are highly dynamic in nature, exhibiting flows, cooling and heating of the local plasma with a wide range of timescales. For accurate values of plasma parameters to be obtained from coronal magneto-seismology this dynamic nature of the plasma needs to be incorporated into the models and the importance of a dynamic background must be assessed.

One feature which should be taken into account when studying the nature of coronal oscillations is the temperature evolution of the coronal plasma. There has been a plethora of observations which show a number of different temperature evolution scenarios (e.g. \opencite{Winebarger03}; \opencite{Nagata}; \opencite{Lopez07}). \inlinecite{Nagata} have observed that there are at least two categories of temperature evolution. Hot loops, which are heated to temperatures $T>2.5$ MK, are seen by the x-ray imagers, e.g., Yohkoh's soft x-ray telescope (SXT) and Hinode's x-ray telescope (XRT). The hot loops are short lived and are seen to undergo relatively fast cooling down to EUV temperatures appearing in EUV imagers, e.g., SOHO/EIT. Another category is cool loops which are observed with temperatures $0.4-1.3$ MK in EUV images and have relatively long lifetimes. The temperature changes on a slow scale when compared to the lifetime of perturbations, or even the loop one's itself and the loops exhibit limited dynamic behaviour. The physical process of the plasma cooling depends upon the loop temperature (cool or hot). It has been found that radiation is the dominant mechanism for the cooling of the EUV loops ($T<2.0$ MK) whereas thermal conduction is the cooling method for loops in the region of $T>2.0$ MK. The majority of observed coronal loops that exhibit oscillations are reported to have a temperature decrease in an exponential form with cooling times of $500-2000$ s \cite{Aschwanden08,Ugarte09}.

Propagating plasma disturbances were detected by SOHO/UVCS along coronal plumes (\opencite{Ofman97}, \citeyear{Ofman99}, \citeyear{Ofman001a}). \inlinecite{Deforest98} observed similar compressive disturbances in polar plumes by using SOHO/EIT. These observed compressive disturbances are interpreted as propagating, slow magneto-acoustic waves, where the suggested damping mechanism for the waves was compressive viscosity (\opencite{Ofman99}, \citeyear{Ofman002b}). This problem was further investigated by \inlinecite{Mendoza04} who considered a gravitational stratification in addition to dissipative processes of thermal conduction, viscosity and radiation. They found that oscillations are even more efficiently damped in stratified plasmas. Similar intensity disturbances were observed in coronal loops by TRACE \cite{Nightingale99,Schrijver99,Moortel00,McEwan06} and EIT/SOHO \cite{Berghmans99}. \inlinecite{Nakariakov00} have also identified these disturbances as slow magneto-acoustic waves and they found that the wave evolution is affected by dissipation and gravitational stratification. Moreover, \inlinecite{Wang03} and \inlinecite{Taroyan07} have observed ($T>6$~MK) hot loop oscillations by SUMER and SXT. They suggested that these oscillations are standing longitudinal (acoustic) waves and are triggered by a footpoint microflare. It was reported that after an initial temperature increase the loop is observed to be cooling.  Recently, \inlinecite{ErdelyiTaroyan08} and \inlinecite{Wang09} have detected 5 minutes quasi-periodic oscillations in transition region and additional five coronal lines by Hinode/EIS at the footpoint of a coronal loop. These oscillations were interpreted as slow magneto-acoustic waves propagating upward from the transition region into the corona. Further, it was found that the amplitude of oscillations decreases with increasing height. It is suggested that the source of these oscillations are the leakage of the photospheric {\it p}-modes through the chromosphere and transition region into the corona, i.e. a similar mechanism that is put forward to explain spicule formation \cite{Pontieu04,Pontieu06} and transition region moss oscillations \cite{Pontieu03}.

A theoretical study of how the slow mode under solar coronal conditions can be extracted from the MHD equations has been done by, e.g. \inlinecite{Roberts06}. The magnetic field was assumed straight in the vertical direction in a gravitationally stratified medium. This theory has been applied to the slow waves in coronal loops observed by SUMER and TRACE. \inlinecite{Luna11} generalised these studies to concurrently occurring magnetic and density stratification, i.e. taking into account the strongly expanding nature of loops that support MHD waves.

The hot loops are thought to be heated by impulsive heating events, e.g. flares, micro-flares, nano-flares. The origin of these energisation mechanisms could be either waves or reconnection (\opencite{Antolin08a}, \citeyear{Antolin08b}, \citeyear{Antolin09}). There have been a number of simulations investigating the evolution of coronal loops after heating events (see, e.g. \opencite{Jakimiec92};  \opencite{Cargill94}; \opencite{Markus09}; \opencite{Taroyan09}, \citeyear{Taroyan10}; \opencite{Taroyanetal10}). Impulsive heating events only last for a relatively short duration and if the plasma reaches a hot enough temperature by the end of the heating, thermal conduction dominates the cooling of the plasma. From calculations by e.g. \inlinecite{Cargill94}, the decrease in temperature due to thermal conduction takes an almost exponential form.

The effect of thermal conduction on MHD waves has been extensively studied. However, all the pervious models have assumed the background plasma is static, i.e. no change in the equilibrium plasma quantities over time. \inlinecite{Ibanez93} studied the propagation of thermal and magnetosonic waves in optically thin plasma. They found that thermal waves are always damped whereas the magnetosonic waves are damped in the range of temperature $10^4$ K $\leq T<10^8$ K due to thermal conduction mechanism. Moreover, \inlinecite{Ofmanwang} and \inlinecite{Mendoza04} have used a 1D nonlinear dissipative MHD model to study the Doppler shift oscillations of hot coronal loops observed by SUMER. The oscillations were interpreted as slow magnetosonic waves. Compressive viscosity was found to be less significant on damping loop oscillations compared to the thermal conduction mechanism. The propagation of MHD slow waves in a static 1D isothermal medium has also been studied by \inlinecite{Moortel03} where thermal conduction and compressive viscosity are considered as damping mechanisms for perturbations. They found that `enhanced' thermal conduction is the dominant damping mechanism of slow waves in coronal loops. Further to this, \inlinecite{Moortel04a} investigate the behaviour of slow MHD waves in a stratified and diverging static atmosphere. The introduction of stratification reduced the effectiveness of thermal conduction on the damping of the slow wave. This result is in disagreement with those of \inlinecite{Mendoza04}, who found an opposite effect for slow standing waves. Further, there is reported a decay of oscillation amplitude by optically thin radiation and area divergence although this is only found to change the rate of damping by a few percent. Overall, it is found in many earlier studies that thermal conduction is an essential cause of the damping of loop oscillations in static equilibria when compared to the other mechanisms under the coronal conditions (i.e. $T=10^6$ K). However, there is an important point to highlight: although thermal conduction seems to be an efficient dissipative mechanism among the various damping mechanisms, considered, it can not efficiently model the {\it observed} decay of amplitudes, when coronal parameters are substituted into the classical formulae of thermal conduction obtained by \inlinecite{Braginskii65} and \inlinecite{Chapman58}, etc. This means that modelling efforts need to consider alternatives. We provide here one.

The damping of oscillations due to a cooling {\it background} coronal plasma is a novel and physically natural idea. \inlinecite{Morton09b} found that the cooling of the background plasma could lead to the damping of even fast kink (i.e. by largely incompressible) oscillations. In addition to this, they noticed that the ratio of fundamental mode to the first harmonic decreases as the loop cools. Furthermore, \inlinecite{Morton09c} found that the damping due to the cooling of the plasma could account for the observed damping in a number of examples of observed transverse oscillations. At the time of writing, this number was almost all the known examples available in the literature (\opencite{Aschwanden99}, \citeyear{Aschwanden02}; \opencite{Nakariakov99}). In a recent theoretical work by \inlinecite{Morton09d}, the effect of a radiative cooling background plasma state on the propagation of magneto-acoustic waves in a uniformly magnetized plasma was investigated. Although the approximation of unbounded uniformity of the plasma may seem to be a severe simplification, this step was necessary in order to give insight into the underlying physics. The radiation mechanism was assumed to be the mechanism for the cooling of the plasma and the plasma cooled exponentially in time, which is capturing well the key features of the observational data. As a result, they found that the slow and fast modes are damped due to cooling. The radiative cooling was shown to damp the slow mode by up to $60\%$ within characteristic lifetimes. This effect is much stronger than the contribution predicted by \inlinecite{Moortel04a} and there was no need of additional, less realistic assumption about the coefficient of thermal conduction.

In this paper, we study further the propagation of longitudinal MHD waves in a uniform magnetized plasma under the influence of a cooling background state. The cooling of the plasma is assumed, here, to be dominated by thermal conduction, so is applicable to oscillations in hot loops. Solutions to the background plasma equations show the cooling profile of the plasma is exponential in time in agreement with observational reports. The dispersion relation which describes the slow modes and their properties is derived. An analytic expression for the time-dependent amplitude of the waves is also derived. The results show that thermal conduction has a dominant effect on the slow modes where their amplitude is strongly damped in plasmas with a dynamic background state. The rate of damping is quantified in the model and is found to be dependent upon the amount of stratification in the plasma and the initial temperature of the background. The damping is also shown to be weakly dependent upon position along the slowly varying equilibrium.
\section{Governing equations}
Consider a homogeneous magnetised plasma where the background temperature is changing as a function of time due to thermal conduction and density is constant in time. The magnetic field is uniform and in the $z$ direction, i.e. $\mathbf{B}_0=B_0\bf{\hat{z}}$.
The governing MHD equations for the background plasma take the following form
\begin{eqnarray}
\frac{\partial{\rho}}{\partial{t}}+\nabla.(\rho\mathbf{v}) =0,\label{eq:cont}\\
\rho\frac{\partial{\mathbf{v}}}{\partial{t}}+\rho(\mathbf{v}.\nabla)\mathbf{v}=-\nabla{p}
+\frac{1}{\mu_0}(\nabla\times\mathbf{B})\times\mathbf{B}+\rho g,\\
\frac{R}{\tilde\mu}\frac{\rho^\gamma}{(\gamma-1)}\left[\frac{\partial{}}{\partial{t}}\frac{T}{\rho^{\gamma-1}}
+(\mathbf{v}.\nabla)\frac{T}{\rho^{\gamma-1}}\right]=\kappa\nabla^2{T},\\
\frac{\partial{\mathbf{B}}}{\partial{t}}=\nabla\times(\mathbf{v}\times\mathbf{B}),\\
{p}=\frac{R}{\tilde\mu}\rho{T},\label{eq:gas-law}
\end{eqnarray}
where $\mathbf{v}$ is the flow velocity, $\mathbf{B}$ is the magnetic field, $g$ is the gravity, $\mu_0$ is the magnetic permeability of free space, $\gamma$ is the ratio of specific heats, $R$ is the gas constant, $\tilde{\mu}$ is the mean molecular weight, $T$ is the temperature, $\kappa\nabla^2{T}$ is thermal conduction term, $\rho$ and $p$ are the plasma density and its pressure, respectively.

 The medium is assumed to be cooling due to thermal conduction with no temporal change in density and it is also assumed that there is no background flow (i.e. $\mathbf{v}_0=0$), so Eqs. $(\ref{eq:cont})-(\ref{eq:gas-law}$) reduce to
\begin{eqnarray}
\mathbf{v}_{0}=0 ,\;\frac{\partial{\rho}_{0}}{\partial{t}}=0,\\ \nabla{p}_{0}=\rho_0g,\label{background_pressure}\\
\rho_{0}=\rho_{0}(z),\\
{p}_{0}=\frac{R}{\tilde\mu}\rho_{0}{T}_{0}, \label{eq: background-gas low} \\
\frac{{R}\,\rho_{0}}{\tilde\mu(\gamma-1)}\frac{\partial{T}_{0}}{\partial{t}}=\kappa\nabla^2T_{0}, \label{eq1}\
\end{eqnarray}
where the (0) index denotes background quantity.

Since Eq. ($\ref{background_pressure}$) can be written as
\begin{equation}
\frac{1}{p_0}dp_0=\frac{l}{H}d\tilde{z},\label{scale_height}
\end{equation}
where $z=l\tilde{z}$ (here $l$ is some characteristic length scale) and $H=p_0/\rho_0g$, and we are interested in a small characteristic scale length which is much smaller than the scale height, $H$, then we arrive at
\begin{equation}
\nabla{p}_{0}\approx0,
\end{equation}
which is representative to a weakly stratified atmosphere, i.e. $l/H\ll1$.
The solution to Eq. ($\ref{eq1}$) by separation of variables, and using Eq. ($\ref{eq: background-gas low}$) gives the temperature as
\begin{equation}
T_{0}(z,t)=T_i\exp{\left(\frac{-(\gamma-1)\lambda\tilde\mu\kappa{t}}{R\rho_i}\right)}[{-c}_{2}z^2+c_{3}z+c_4],
\end{equation}
and the density as
\begin{equation}
\rho_0(z)=\frac{\rho_i}{{-c}_{2}z^2+c_{3}z+c_4}, \label{back_den}
\end{equation}
which is physically valid when $c_2, c_3\ll1$ and $c_3>c_2$, where $\lambda$ is the separation constant, $c_2=\lambda/2$, $c_3$ and $c_4$ are constants, $T_i$ is the initial temperature at $z=0$ and $t=0$, and $\rho_i$ is the density at $z=0$.

Note that the background temperature decreases exponentially with time. It has been shown for radiative cooling loops that an exponential profile provides a good fit for the observed cooling (\opencite{Aschwanden08}; \opencite{Morton09b}, \citeyear{Morton09c}). To the best of our knowledge, there are no such observations for hot loops cooling due to thermal conduction, that is why $\partial\rho_0/\partial t=0$ here.

The form of the density profile in Eq. ($\ref{back_den}$) is consistent with a weakly stratified atmosphere. For a gravitational stratified atmosphere the density profile would be formed to be
\begin{equation}
\rho_0=\rho_i\exp\left(-\frac{z}{H}\right).\label{grav-stratif}
\end{equation}
The hydrostatic scale height $H$ is given by $H=47T\left[{\frac{\textmd{Mm}}{\textmd{MK}}}\right]$, hence for hot loops $H$ is in the range $94-282$ Mm. 
In the case of weak stratification, i.e. $l/H\ll1$, the density can be approximated as
\begin{equation}
\rho_0\approx\rho_i\left(1-\frac{z}{H}+\frac{1}{2}\left(\frac{z}{H}\right)^2\right).\label{weak-stratif}
\end{equation}
Eq. ($\ref{back_den}$) can be written
$$
\rho_0=\frac{\rho_i}{1+y}\approx\rho_i(1-y)\simeq\rho_i\left[1-(-c_2z^2+c_3z+c_4-1)\right],
$$
upon the assumption $y$ is small. Comparing to Eq. ($\ref{weak-stratif}$), then the values for the constants are $c_4=1$, $c_3=\frac{1}{H}$ and $c_2=\frac{\lambda}{2}=\frac{1}{2}\left(\frac{1}{H}\right)^2$ and hence $y$ is small.

Perturbing the background equations, the linearized MHD equations can be found by writing all variables in the form
$$
f(z,t)=f_0(z,t)+f_1(z,t),
$$
and neglecting all terms containing squared perturbed variables. Here the subscript (0) represents the equilibrium quantities and the subscript (1) indicates the perturbed quantities.

 Since thermal conduction is known to have a strong effect on slow modes, we will concentrate our analysis on the properties of slow modes. It has been shown by a number of authors, e.g. \inlinecite{Roberts06} and \inlinecite{Luna11}, that the slow modes can be isolated by assuming $v_{1x} = v_{1y} = 0$. Therefore the linear dissipative MHD equations for the slow modes in the presence of thermal conduction reduce to a 1-D system given by
\begin{eqnarray}
\frac{\partial\rho_{1}}{\partial{t}}+\rho_{0}\frac{\partial{v_{1}}}{\partial z}+v_{1}\frac{\partial\rho_{0}}{\partial{z}}=0,\label{Eq:dimless-cont}\\
\rho_{0}\frac{\partial{v}_{1}}{\partial{t}}=-\frac{\partial{p}_{1}}{\partial z},\label{Eq:motion}\\
\frac{R}{\tilde\mu}\left[\frac{\rho_{1}}{\gamma-1}\frac{\partial{T}_{0}}{\partial{t}}+\frac{\rho_{0}}{\gamma-1}\frac{\partial{T}_{1}}{\partial{t}}+\rho_{0}T_{0}\frac{\partial{v}_{1}}{\partial z}\
-\frac{T_0}{\gamma-1}{v}_{1}\frac{\partial\rho_{0}}{\partial z}\right]=\kappa\nabla^2T_{1},\\
\frac{\partial\mathbf{B}_{1}}{\partial{t}}=\nabla\times(\mathbf{v}_{1}\times\mathbf{B}_{0})=B_{0}\frac{\partial\mathbf{v}_{1}}{\partial{z}}-\mathbf{B}_{0}\
(\frac{\partial\mathbf{v}_{1}}{\partial z}),\\
{p}_{1}=\frac{R}{\tilde{\mu}}\left[\rho_{0}T_{1}+\rho_{1}T_{0}\right],\\
\nabla\cdot\mathbf{B}_{1}=0.\label{Eq:magnetic-field}
\end{eqnarray}
Here $v_1\equiv v_{1z}$.
Dimensionless variables can now be introduced to simplify the equations, where the following dimensionless quantities are suggested:
\begin{eqnarray}
\tilde\rho_{1}=\frac{\rho_{1}}{\rho_{i}},\,\tilde\rho_{0}=\frac{\rho_{0}}{\rho_{i}},\;\tilde{p}_{1}=\frac{p_{1}}{p_{i}},\ \tilde{T}_{1}=\frac{T_{1}}{T_{i}},\,\tilde{T}_{0}=\frac{T_{0}}{T_{i}},\,
\mathbf{\tilde{v}}_{1}=\frac{\mathbf{v}_{1}}{c_{si}},\,\\\nonumber c_{si}=\frac{l}{\tau},\,
c^2_{si}=\frac{\gamma{p}_{i}}{\rho_{i}},\,\tilde{t}=\frac{t}{\tau},\,\tilde{z}=\frac{z}{l},\,\tilde\lambda=\left(\frac{l}{H}\right)^2\!\!,\,
\tilde{c_2}=lc_2,\,\tilde{c_3}=lc_3.
\end{eqnarray}
Here $c_{si}$ is the initial sound speed, $l$ is the wavelength of the oscillations and $\tau$ is the sound travel time of a wavelength.
Thus the equation of energy in terms of dimensionless variables removing tilde, will be
\begin{eqnarray}
\rho_{1}\frac{\partial{T}_{0}}{\partial{t}}+\rho_{0}\frac{\partial{T}_{1}}{\partial{t}}+(\gamma-1)\rho_{0}{T}_{0}\nabla.\mathbf{v}_{1}\
-T_{0}v_{1}\frac{\partial\rho_{0}}{\partial{z}}=d\frac{\partial^2T_{1}}{\partial{z}^2}, \
\;\textmd{where}\quad d=\frac{(\gamma-1)\kappa\,\rho_i{T_i}}{\gamma\,{p}_i^2\,\tau}.\label{Eq:dimless-energy1}
\end{eqnarray}
Using coronal values, (see e.g. \opencite{Moortel03}), we find $d$ is a small quantity, where the standard coronal values of all variables
\begin{equation}
\left\{
\begin{array}{ll}
T_0=1-6\;\textmd{MK},\\
\rho_0=1.67\times10^{-12}\; \textmd{kg m}^{-3},\\
\kappa=10^{-11}\,T_0^{5/2}\; \textmd{W m}^{-1}\; \textmd{deg}^{-1},\\
\tilde{\mu}=0.6,\\
R=8.3\times10^3\;\textmd{m}^2\;\textmd{s}^{-2}\;\textmd{deg}^{-1},\\
\gamma=5/3,\\
\tau=300\; \textmd{s},\\
\end{array}
\right.
\end{equation}
give a value of $d=0.04$ for $T=1$ MK and $d=0.61$ for $T=6$ MK.

From the linearized ideal gas law equation
\begin{equation}
{p}_{1}=\rho_{0}{T}_{1}+\rho_{1}{T}_{0},\label{eq2}\
\end{equation}
and the continuity equation
\begin{equation}
\frac{\partial\rho_{1}}{\partial{t}}+\rho_{0}\frac{\partial{v}_{1}}{\partial z}+v_{1}\frac{\partial\rho_{0}}{\partial{z}}=0,\label{eq:dimless-cont}
\end{equation}
the energy equation takes the following form
\begin{equation}
\frac{\partial{p}_{1}}{\partial{t}}+\gamma\rho_{0}{T}_{0}\frac{\partial{v}_{1}}{\partial{z}}
=d\frac{\partial^2T_{1}}{\partial{z}^2}.\label{eq:dimless-energy}
\end{equation}
Differentiating Eq. ($\ref{eq:dimless-energy}$) with respect to $z$ and using Eq. ($\ref{Eq:motion}$), we arrive at
\begin{equation}
\frac{\partial^2{v}_{1}}{\partial{t}^2}-T_{0}\frac{\partial^2{v}_{1}}{\partial{z}^2}\
=-\frac{d}{\gamma\rho_{0}}\frac{\partial^3{T}_{1}}{\partial{z}^3}.\label{Eq:modified-dimless-energy}
\end{equation}
To solve the equation of energy, we start from the continuity equation and using the gas law equation to find an equation in terms of velocity variable.

Substituting Eq. ($\ref{eq2}$) into Eq. ($\ref{eq:dimless-cont}$), we obtain
\begin{equation}
\frac{1}{T_{0}}\frac{\partial{p}_{1}}{\partial{t}}-\frac{p_{1}}{T_{0}^2}\frac{\partial{T}_{0}}{\partial{t}}\
-\frac{\rho_{0}}{T_{0}}\frac{\partial{T}_{1}}{\partial{t}}+\frac{\rho_{0}T_{1}}{T_{0}^2}\frac{\partial{T}_{0}}{\partial{t}}\
+\rho_{0}\frac{\partial{v}_{1}}{\partial{z}}+{v}_{1}\frac{\partial\rho_{0}}{\partial{z}}=0.\label{eq3}\\
\end{equation}
The dimensionless background temperature is
$$
T_0(z,\,t)=\frac{\exp(-\lambda{d}{t})}{\rho_0(z)},
$$
so
\begin{equation}
\frac{\partial{T}_{0}}{\partial{t}}=\delta{T}_{0},\;\ \textmd{where}\quad\delta=-\lambda{d}.\label{eq4}\
\end{equation}
Then, after substituting ($\ref{eq4}$) and multiplying by ${T}_{0}$, Eq. ($\ref{eq3}$) will be
\begin{equation}
\frac{\partial{p}_{1}}{\partial{t}}-\delta{p}_{1}-\rho_{0}\frac{\partial{T}_{1}}{\partial{t}}+\delta\rho_{0}T_{1}\
+\rho_{0}T_{0}\frac{\partial{v}_{1}}{\partial{z}}+T_{0}{v}_{1}\frac{\partial\rho_{0}}{\partial{z}}=0.\label{Eq:cont-gas_law}
\end{equation}
Substituting this equation into Eq. $(\ref{eq:dimless-energy})$, we obtain the following equation
\begin{equation}
(\gamma-1)\rho_0T_0\frac{\partial^2v_1}{\partial t \partial z}-\delta\rho_0T_0\frac{\partial v_1}{\partial z}-\frac{\partial}{\partial t}(T_0v_1)\frac{\partial\rho_0}{\partial z}=d\frac{\partial^3T_1}{\partial t\partial z^2}-\rho_0\frac{\partial^2T_1}{\partial t^2}+\delta\rho_0\frac{\partial T_1}{\partial t},
\end{equation}
which represents another relation between the perturbed temperature and perturbed velocity in addition to Eq. ($\ref{Eq:modified-dimless-energy}$). Now, we aim to reach the governing equation for the velocity perturbation. Differentiating Eq. $(\ref{Eq:cont-gas_law})$ once with respect to $t$ and three times with respect to $z$, and using Eqs. ($\ref{eq:dimless-energy}$) and ($\ref{Eq:modified-dimless-energy}$) with some cumbersome algebraic operations, we arrive at
\begin{eqnarray}
\gamma\frac{\rho_{0}}{d}\frac{\partial^4{v}_{1}}{\partial{t}^4}-\gamma\left[\frac{\rho_{0}\delta}{d}+\frac{3}{\rho_0^2}
\left(\frac{\partial\rho_0}{\partial{z}}\right)^2-\frac{2}{\rho_0}\frac{\partial^2\rho_{0}}{\partial{z}^2}
-\frac{1}{\rho_{0}}\frac{\partial\rho_0}{\partial{z}}\frac{\partial}{\partial{z}}+\frac{\partial^2}{\partial{z}^2}\right]
\frac{\partial^3{v}_{1}}{\partial{t}^3}\nonumber\\
+\,\gamma\left[\frac{3\delta}{\rho_0^2}\left(\frac{\partial\rho_0}{\partial{z}}\right)^2-\frac{2\delta}{\rho_0}
\frac{\partial^2\rho_{0}}{\partial{z}^2}-\frac{\delta}{\rho_{0}}\frac{\partial\rho_0}{\partial{z}}\frac{\partial}{\partial{z}}
+(\delta-\frac{\rho_{0}T_{0}}{d})\frac{\partial^2}{\partial{z}^2}\right]\frac{\partial^2{v}_{1}}{\partial{t}^2}\nonumber\\
-\left[\frac{\gamma\delta\rho_{0}{T}_{0}}{d}\frac{\partial^2{}}{\partial{z}^2}-2\frac{\partial{T}_{0}}{\partial{z}}
\frac{\partial^3}{\partial{z}^3}-T_{0}\frac{\partial^4}{\partial{z}^4}\right]\frac{\partial{v}_{1}}{\partial{t}}
+\left[2\delta\frac{\partial{T}_{0}}{\partial{z}}\frac{\partial^3}{\partial{z}^3}
+\delta{T}_{0}\frac{\partial^4{}}{\partial{z}^4}\right]{v}_{1}=0.\label{eq5}
\end{eqnarray}
To simplify Eq. ($\ref{eq5}$) we neglect all terms of order $d\delta$ and higher degree in $d$ or $\delta$ as $d$ and $\delta$ are small parameters. Then, after multiplying equation ($\ref{eq5}$) by $d$,  we obtain the governing equation
\begin{eqnarray}
\gamma\rho_{0}\frac{\partial^4{v}_{1}}{\partial{t}^4}-\gamma\rho_{0}T_{0}\frac{\partial^2}{\partial{z}^2}\frac{\partial^2{v}_{1}}{\partial{t}^2}=
\gamma\left[\rho_{0}\delta+\frac{3d}{\rho_0^2}
\left(\frac{\partial\rho_0}{\partial{z}}\right)^2-\frac{2d}{\rho_0}\frac{\partial^2\rho_{0}}{\partial{z}^2}\right.\nonumber \\
\left.-\frac{d}{\rho_{0}}
\frac{\partial\rho_0}{\partial{z}}\frac{\partial}{\partial{z}}+d\frac{\partial^2}{\partial{z}^2}\right]
\frac{\partial^3{v}_{1}}{\partial{t}^3}
+\left[\gamma\delta\rho_{0}{T}_{0}\frac{\partial^2}{\partial{z}^2}-2d\frac{\partial{T}_0}{\partial{z}}
\frac{\partial^3}{\partial{z}^3}-dT_{0}\frac{\partial^4{}}{\partial{z}^4}\right]\frac{\partial{v}_{1}}{\partial{t}}.
\end{eqnarray}
The background quantities are slowly varying when compared to the perturbed quantities, so eliminating the small terms gives
\begin{equation}
\frac{\partial}{\partial t}\left[\frac{\partial^2{v}_{1}}{\partial{t}^2}-T_{0}\frac{\partial^2v_1}{\partial{z}^2}\right]=
\delta\left[\frac{\partial^2v_1}{\partial{t}^2}+T_0\frac{\partial^2v_1}{\partial{z}^2}\right]+\frac{d}{\rho_0}\frac{\partial^2}{\partial z^2}\left[\frac{\partial^2{v}_{1}}{\partial{t}^2}-\frac{T_0}{\gamma}\frac{\partial^2v_1}{\partial{z}^2}\right].\label{eneq}
\end{equation}
In the case of no thermal conduction, Eq. ($\ref{eneq}$) reduces to the wave equation that represents (longitudinal) acoustic mode propagating in a flux tube and has the form
\begin{equation}
\frac{\partial^2v_1}{\partial t^2}-T_0\frac{\partial^2 v_1}{\partial z^2}=0,\label{wave}
\end{equation}
while in the case of unstratified atmosphere, i.e. no change in the background plasma quantities, Eq. ($\ref{eneq}$) leads to the model governing equation found by \inlinecite{Moortel03} as follows
\begin{equation}
\frac{\partial}{\partial t}\left[\frac{\partial^2{v}_{1}}{\partial{t}^2}-T_{0}\frac{\partial^2v_1}{\partial{z}^2}\right]=d\,\frac{\partial^2}{\partial z^2}\left[\frac{\partial^2{v}_{1}}{\partial{t}^2}-\frac{1}{\gamma}\frac{\partial^2v_1}{\partial{z}^2}\right].
\end{equation}
Eq. ($\ref{eneq}$) can now be checked. Assuming the background is constant in time and space, i.e. static and unstratified, one can Fourier-analyse all perturbations, $\exp({i}(\omega{t}-{k}{z}))$. Hence, Eq. ($\ref{eneq}$) reduces to the dispersion relation
\begin{equation}
\omega^3-id\,\omega^2{k}^2-\omega{k}^2+i\frac{d}{\gamma}{k}^4=0,
\end{equation}
where $\omega$ is the frequency and $k$ is the wavenumber. This dispersion relation has been obtained first by \inlinecite{Field65}. This provides confidence that Eq. ($\ref{eneq}$) is consistent with earlier studies (see, e.g. \opencite{Field65} and \opencite{ Moortel03}).
Note that, in the limit ${d}\longrightarrow0$, i.e. no thermal conduction, Eq. ($\ref{eneq}$) reduces to Eq. ($\ref{wave}$) i.e. $\omega^2=k^2\Longrightarrow\omega=\pm{k}$ which is, rightly, the undamped (longitudinal) acoustic mode.
\section{Analytical solutions}
We now seek to find an analytic solution to the governing Eq. ($\ref{eneq}$). Unlike previous models which include thermal conduction, there is now a time dependence due to the temporally changing background temperature. This means we can not Fourier analyse in time. Instead, since the RHS of Eq. ($\ref{eneq}$) has derivatives multiplied by a small factor, $d$, the WKB approximation (see, e.g. \opencite{Bender}) can be applied to find an approximate solution. The WKB approximation has a greater accuracy for the smaller value of $d$, however, it can still provide an accurate solution even for $d\approx1$. Since the variables in Eq. ($\ref{eneq}$) depend on time $t$ and space $z$, we introduce `slow' variable $t_1=dt$ and `local' variable $\zeta=dz$ to solve this equation. The slow timescale physically means that the conductive cooling timescale is longer than the period of the oscillations. The `local' lengthscale means any significant changes in quantities in the $z$ direction occur on length scales longer than the wavelength of the oscillation. The velocity in terms of the new scaled variables and by the WKB approximation has the form
\begin{equation}
v_1(\zeta,\,t_1)=Q(\zeta,\,t_1)\exp\left(\frac{i}{d}\Theta(\zeta,\,t_1)\right),\label{wkb}
\end{equation}
where $Q(\zeta,\,t_1)$ and $\Theta(\zeta,\,t_1)$ are functions to be calculated.

Substituting Eq. ($\ref{wkb}$) into Eq. ($\ref{eneq}$) and taking the largest order terms in $d$, which is $d^{-4}$, we obtain
\begin{equation}
\left(\frac{\partial\Theta}{\partial{t}_1}\right)^2-c_s^2(z,\,t)\left(\frac{\partial\Theta}{\partial\zeta}\right)^2=0,
\qquad c_s(z,\,t)=\sqrt{T_0}.\label{freqeq}
\end{equation}
If we assume $\partial\Theta/\partial t=\omega(z,\,t)$ and $\partial\Theta/\partial\zeta=k(z,\,t)$, where $\omega(z,\,t)$ is frequency and $k(z,\,t)$ is wavenumber, then Eq. ($\ref{freqeq}$) is in fact a temporally and spatially dependent dispersion relation for the longitudinal (acoustic) mode.

The next largest order terms in $d$ (of order $d^{-3}$) give the equation for the amplitude,
\begin{equation}
-2\gamma\rho_0c_s\,\frac{\partial{Q}}{\partial{t}_1}\frac{\partial\Theta}{\partial\zeta}
+\frac{\gamma}{2}\lambda\rho_0c_s\,Q\frac{\partial\Theta}{\partial\zeta}
-\gamma\rho_0c_s\, \frac{\partial{c}_s}{\partial\zeta}Q\frac{\partial\Theta}{\partial\zeta}-(\gamma-1)c_s\,Q
\left(\frac{\partial\Theta}{\partial\zeta}\right)^3 +2\gamma\rho_0c_s^2\frac{\partial{Q}}{\partial\zeta}\frac{\partial\Theta}{\partial\zeta}=0,
\end{equation}
which, after some algebra, has the form
\begin{equation}
\frac{-1}{c_s}\,\frac{\partial{Q}}{\partial{t}_1}+\frac{\partial{Q}}{\partial\zeta}
+\left[\frac{\lambda}{4c_s}-\frac{1}{2c_s}\, \frac{\partial{c}_s}{\partial\zeta}-\frac{(\gamma-1)}{2\gamma\rho_0c_s}
\left(\frac{\partial\Theta}{\partial\zeta}\right)^2\right]Q=0. \label{ampliteq}
\end{equation}
Eqs. ($\ref{freqeq}$) and ($\ref{ampliteq}$) will be solved by using the method of characteristics so we need to derive boundary conditions at $z=0$.

To achieve this we study a thin layer around $z=0$ where we may assume the spatial gradients of both $T_0$ and $\rho_0$ are very small, so they can be considered constant in space in this region. This enables the use of Fourier-analysis such that, there is $\sim\! \exp(ikz)$, for perturbed variables. Eq. ($\ref{eneq}$) reduces to
\begin{equation}
\gamma\rho_0\frac{d^4v_1}{d{t}^4}+\gamma[\rho_0\lambda d+dk^2]\frac{d^3v_1}{d{t}^3}
+\gamma\rho_0T_0k^2\frac{d^2v_1}{d{t}^2}-[\gamma\lambda{d}\rho_0T_0k^2-dT_0k^4]\frac{d{v_1}}
{d{t}}=0.\label{eq:Ene-charact}
\end{equation}
Next, we apply the WKB method to this equation by assuming that the perturbation in terms of the variable $t_1$, where $t_1=dt$, has the form
\begin{equation}
v_1(t_1)=Q_1(t_1)\exp\left(\frac{i}{d}\Theta_1(t_1)\right).
\end{equation}
Substituting into Eq. ($\ref{eq:Ene-charact}$), the highest order equation in $d$ gives
\begin{equation}
\frac{d\Theta_1}{d{t}_1}=c_s(t_1)k.\label{Eq:Bound-con1}
\end{equation}
Eq. ($\ref{Eq:Bound-con1}$) has the solution
\begin{equation}
\Theta_1(t_1)=\frac{2k}{\lambda}[1-c_s(t_1)].\label{Bound-con1}
\end{equation}
The next order equation in $d$ is
\begin{equation}
\frac{d{Q_1}}{d{t}_1}-\left(\frac{\lambda}{4}-\frac{(\gamma-1)}{\gamma}\frac{k^2}{2}\right)Q_1=0,
\end{equation}
which has the solution
\begin{equation}
Q_1(t_1)=\exp\left[\left(\frac{\lambda}{4}-\frac{(\gamma-1)}{\gamma}\frac{k^2}{2}\right)t_1\right],\label{Bound-con2}
\end{equation}
where $k$ is the wavenumber at $z=0$.

We now have the necessary approximate boundary conditions at $z=0$ to proceed with the solutions of Eqs. ($\ref{freqeq}$) and ($\ref{ampliteq}$) using the method of characteristics.

Then, the solutions of Eqs. ($\ref{freqeq}$) and ($\ref{ampliteq}$) can be found by the method of characteristics and we introduce the variables $r$ and $s$ to parameterise these equations. The characteristic equations of Eq. ($\ref{freqeq}$) are given by
$$
\frac{\partial{t_1}}{\partial{s}}=\frac{-1}{c_s},\qquad \frac{\partial\zeta}{\partial{s}}=1, \qquad \frac{\partial\Theta}{\partial{s}}=0,
$$
with boundary conditions for the characteristics curves
$$
t_1(r,0)=r, \qquad \zeta(r,0)=0,\qquad \Theta(r,0)=\Theta_1(r),
$$
where $\Theta_1(r)$ is given by ($\ref{Bound-con1}$) and $(r,0,\Theta_1)$ is a point in the curve $(t_1(r,s),\zeta(r,s),\Theta(r,s))$ which lies in the surface $\{(t_1,\zeta,\Theta(\zeta,t_1))\}$ that represents the graph of the function $\Theta(\zeta,t_1)$.
The solutions of the characteristic equations are
\begin{equation}
s(\zeta,t_1)=\zeta,\label{Eq:parameter-s}
\end{equation}
\begin{equation}
r(\zeta,t_1)=\frac{-2}{\lambda}\,\ln\left[\exp(\frac{-\lambda{t_1}}{2})-\frac{\lambda{d}}{2\sqrt{c_2}}\,
\arctan\left(\frac{c_3\sqrt{F(\zeta)}+2c_2(\zeta-\frac{c_3d}{2c_2})}{2\sqrt{c_2}\sqrt{F(\zeta)}-c_3\sqrt{c_2}(\zeta
-\frac{c_3d}{2c_2})}\right)\right],\label{Eq:parameter-r}
\end{equation}
\begin{equation}
\Theta(\zeta,t_1)=\frac{-2k}{\lambda}\left(\exp(\frac{-\lambda{t_1}}{2})-\frac{\lambda{d}}{2\sqrt{c_2}}\,
\arctan\left(\frac{c_3\sqrt{F(\zeta)}+2c_2(\zeta-\frac{c_3d}{2c_2})}{2\sqrt{c_2}\sqrt{F(\zeta)}-c_3\sqrt{c_2}(\zeta
-\frac{c_3d}{2c_2})}\right)-1\right).\label{Eq:frequency}
\end{equation}
Similarly, the characteristic equations of Eq. ($\ref{ampliteq}$) are
$$
\frac{\partial{t_1}}{\partial{s}}=\frac{-1}{c_s},\qquad \frac{\partial\zeta}{\partial{s}}=1, \qquad \frac{\partial Q}{\partial{s}}=\left(\frac{\lambda}{4c_s}-\frac{1}{2c_s}\, \frac{\partial{c}_s}{\partial\zeta}-\frac{(\gamma-1)}{2\gamma\rho_0c_s}
\left(\frac{\partial\Theta}{\partial\zeta}\right)^2\right)Q,
$$
with boundary conditions
$$
t_1(r,0)=r, \qquad \zeta(r,0)=0,\qquad Q(r,0)=Q_1(r),
$$
where $Q_1(r)$ is given by ($\ref{Bound-con2}$) and $(r,0,Q_1)$ is a point in the curve $(t_1(r,s),\zeta(r,s),Q(r,s))$ which lies in the surface $\{(t_1,\zeta,Q(\zeta,t_1))\}$ that represents the graph of the function $Q(\zeta,t_1)$.
The solutions of the characteristic equations are Eqs. ($\ref{Eq:parameter-s}$), ($\ref{Eq:parameter-r}$) and
\begin{eqnarray}
Q(\zeta,t_1)=\exp\left[\left(\frac{\lambda}{4}-\frac{\gamma-1}{\gamma}\frac{k^2}{2}\right)\left\{r+\frac{d}{\sqrt{c_2}}
\exp(\frac{\lambda{t_1}}{2})\arctan\left(\frac{c_3\sqrt{F(\zeta)}+2c_2(\zeta-\frac{c_3d}{2c_2})}{2\sqrt{c_2}\sqrt{F(\zeta)}-c_3\sqrt{c_2}(\zeta
-\frac{c_3d}{2c_2})}\right)\right\}\right.\nonumber \\ \label{Eq:amplitude}
\left.+\,\frac{1}{4}\ln\left(\frac{d^2}{F(\zeta)}\right)\right],
\end{eqnarray}
where
\begin{equation}
F(\zeta)=-c_2\zeta^2+c_3d\zeta+d^2=\frac{d^2}{\rho_0(\zeta)}.\label{Eq:density-d_parameter}
\end{equation}
Taking Taylor expansion of the amplitude and ignoring the small terms under the assumption of slowly varying background coronal plasma and weakly stratified atmosphere, we arrive at
\begin{equation}
Q(\zeta,t_1)=1-\left(\frac{\gamma-1}{\gamma}\frac{k^2}{2}-\frac{\lambda}{4}\right)t_1-\frac{1}{4}\sqrt\lambda\;\zeta.
\end{equation}
Eq. ($\ref{wkb}$), combined with Eqs. $($\ref{Eq:parameter-r}$)-($\ref{Eq:density-d_parameter}$)$, provides a full solution to Eq. ($\ref{eneq}$). All other perturbed quantities can now be calculated using Eqs. $($\ref{Eq:dimless-cont}$)-($\ref{Eq:magnetic-field}$)$. The equations may not reveal much information yet as they are very complicated in nature. We examine a limiting case for the amplitude to reveal some properties. Assume that $\lambda=0$ (unstratified atmosphere), then the system of equations  ($\ref{Eq:parameter-r}$), ($\ref{Eq:amplitude}$) and ($\ref{Eq:density-d_parameter}$) reduces to
\begin{equation}
Q(\zeta,t_1)=\exp\left[\left(-\frac{(\gamma-1)}{\gamma}\frac{k^2}{2}\right) t_1\right].
\end{equation}
 This limit is in good agreement with its counterpart in \inlinecite{Moortel03}. The amplitude of the wave is damped and the damping is dependent upon the measure of $d$ and $k$.
\section{Numerical calculations}
It has been shown in \inlinecite{Morton09d} that the WKB estimates provide good approximations to the frequency and amplitude variations in time and space when the plasma is cooling due to radiation. We see no reasons why this should be any different here, where thermal dissipation in the format of thermal conduction is considered. Numerical calculations are now used to evaluate how the analytic solutions of the MHD waves behave in a system with variable background.

\begin{figure}[h]
\vspace{-1 cm}
\centering
\hspace{-0.8 cm}
\subfloat{\label{Amp-damping1a}\includegraphics[height=0.4\textheight, width=0.375\textwidth]{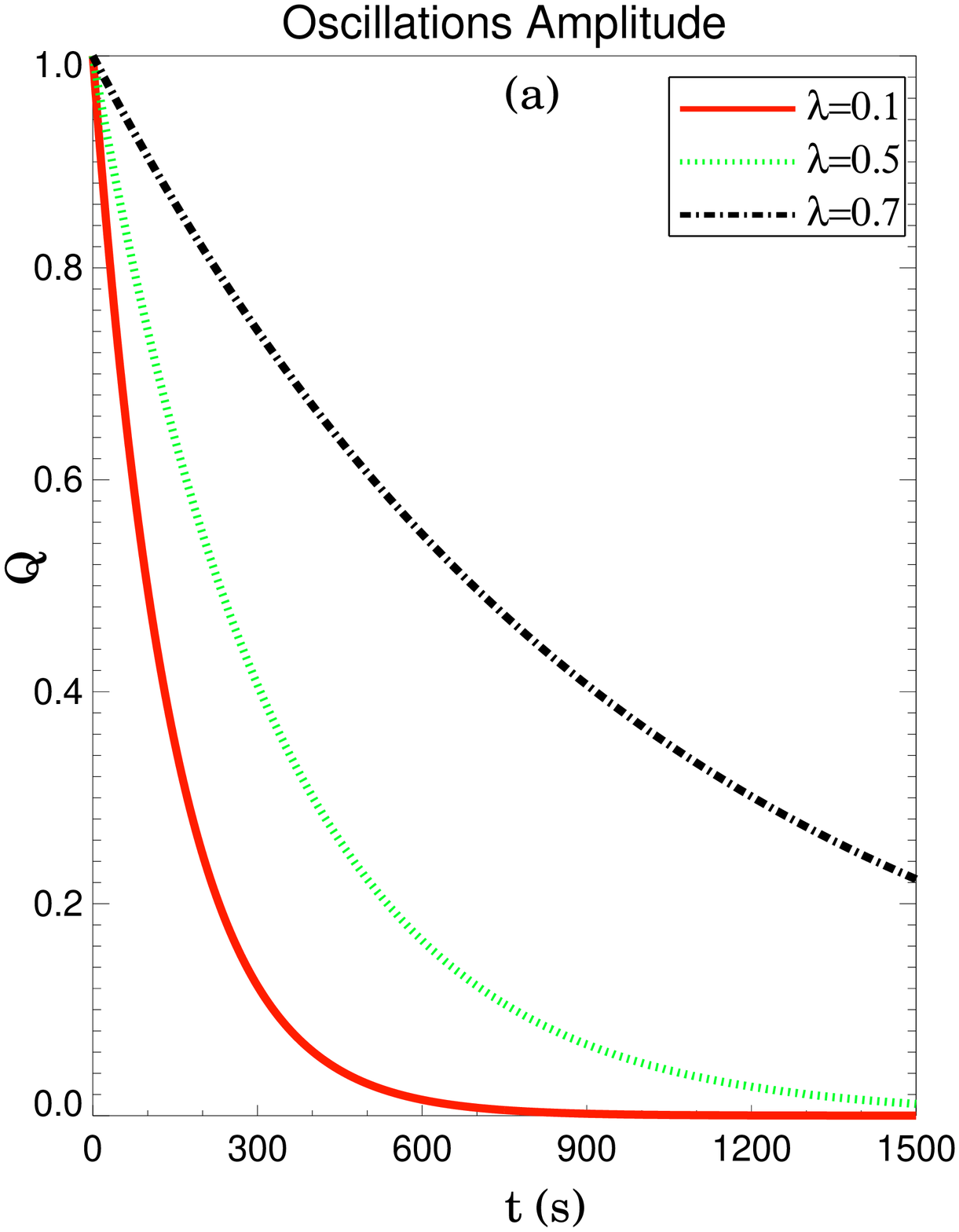}}
\hspace{-0.9 cm}
\subfloat{\label{Amp-damping1b}\includegraphics[height=0.4\textheight, width=0.375\textwidth]{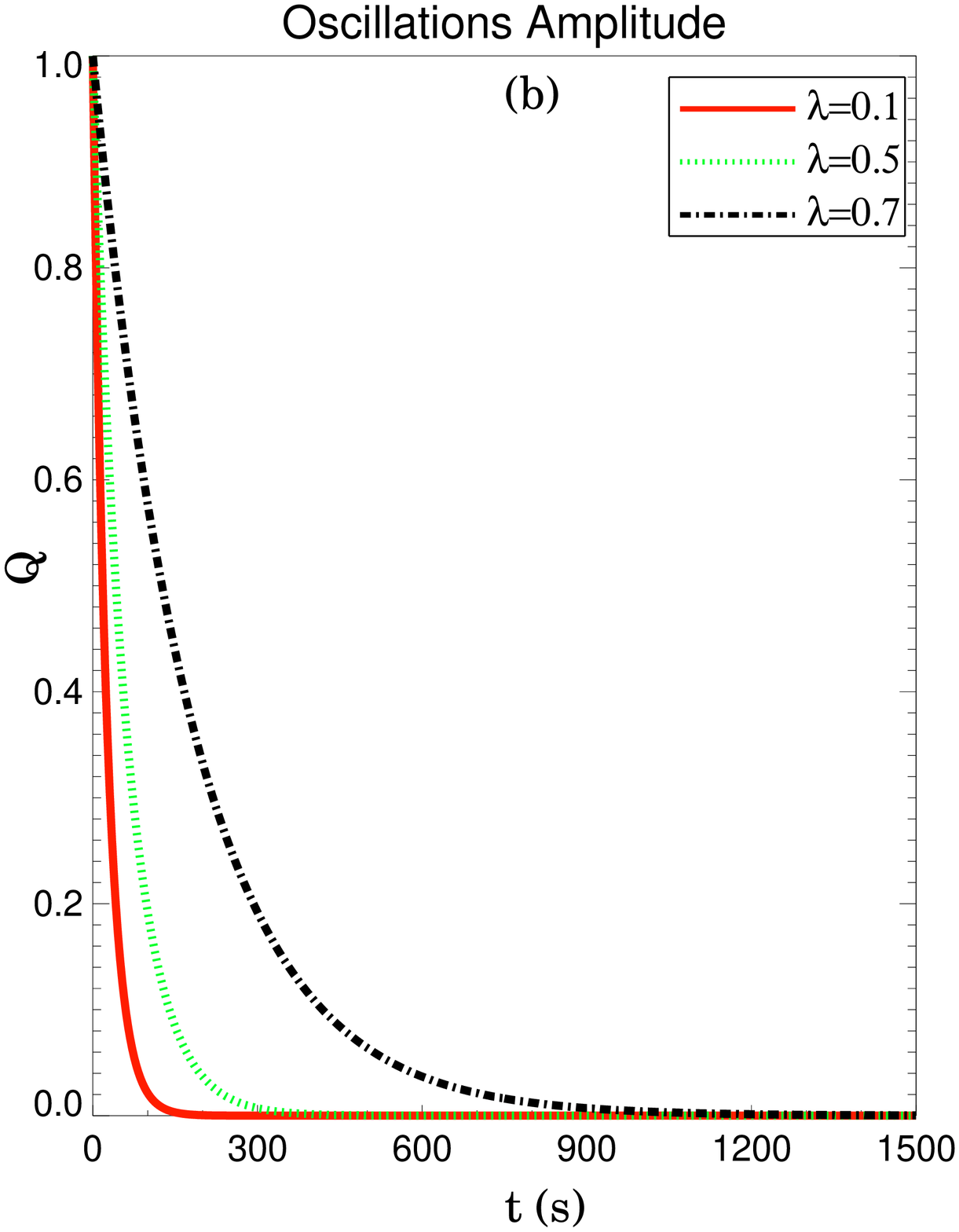}}
\hspace{-0.9 cm}
\subfloat{\label{Amp-damping1c}\includegraphics[height=0.4\textheight, width=0.375\textwidth]{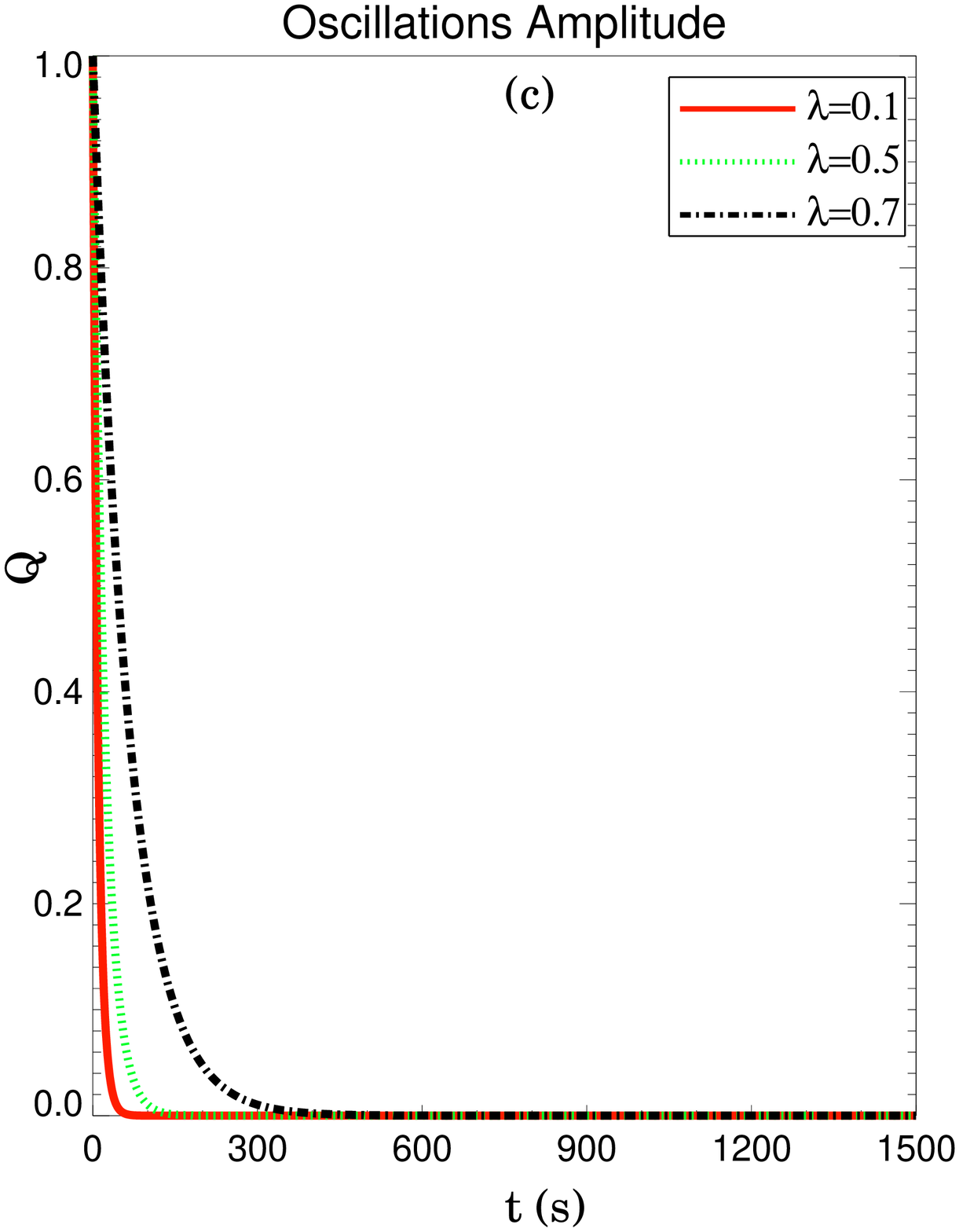}}
\hspace{-0.8cm}
\vspace{-1 cm}
\caption{The graphs show the amplitude of oscillations with different values of $\lambda$ $(0.1, 0.5, 0.7)$ characterising stratification and specific value of $d$, i.e. the value of thermal ratio, at $z=0$. (a) $d=0.04$ ($T=1$ MK), (b) $d=0.22$ ($T=3$ MK),  (c) $d=0.61$ ($T=6$ MK).}\label{Amp-damping}
\end{figure}

The amplitude of slow wave is computed using characteristic coronal values. In Figure \ref{Amp-damping} we plot the amplitude of the longitudinal (acoustic) wave for different values of $\lambda$ (which is the separation constant and is related to the squared reciprocal of scale height) and the value of thermal ratio, $d$, at $z=0$. It is found that the amplitude decays rapidly in time for all values of $\lambda$ and $d$, over typical observed timescales of oscillations. In particular, Figure \ref{Amp-damping1a} depicts that the oscillation amplitude declines dramatically under the coronal values of temperature $10^6$ K. Figure \ref{Amp-damping1b} illustrates a faster decay of the wave amplitude in a region of temperature $3\times10^6$ K, i.e. for even hotter loops. The decay of the amplitude is enhanced even further for very hot (SXT/XRT) loops, i.e. $T=6\times10^6$ K as evidenced by Figure \ref{Amp-damping1c}. In each plot shown in Figure \ref{Amp-damping}, it is also clearly seen that increasing the stratification of the plasma, i.e. increasing $\lambda$, decreases the rate of the damping. This is consistent with \inlinecite{Moortel04a}, who showed that adding stratification decreased the rate of damping of the slow mode due to thermal conduction. The new and important result here is that the rate of damping is much stronger than that predicted by \inlinecite{Moortel04a}. Further, there is no need here to artificially amplify the heat conduction as opposed to \inlinecite{Moortel04a}. A note of caution is, however, needed when interpreting these results. The temperature of the plasma drops rapidly due to the cooling. As $T\rightarrow2$ MK radiation will start to become the dominant damping mechanism. So as $T$ reaches 2 MK the damping of the longitudinal mode may not look as shown here. A full numerical simulation incorporating the variable background and the variable value of $d$ will need to undertaken that is well beyond the scope of the present study and may require a full numerical approach even for an initial insight because of the far to complex mathematical modelling (e.g. generating background flows, etc.).

\begin{figure}[!h]
\centering
\includegraphics[height=0.5\textheight,width=0.5\textwidth]{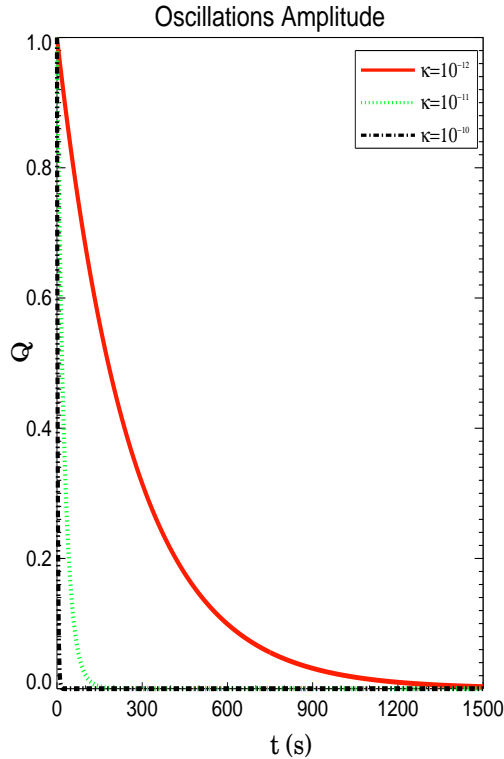}
\vspace{-1 cm}
\caption{The graph shows the amplitude of oscillations with different values of the thermal conduction coefficient, $\kappa=(10^{-10},10^{-11},10^{-12})$ at $z=0$ and $\lambda=0.1$ where $T=3$ MK.}\label{Amp-damping_k}
\end{figure}

In Figure \ref{Amp-damping_k} we show how the magnitude of the thermal conduction coefficient, $\kappa$, effects the rate of damping of the longitudinal (acoustic) mode. Altering the value of $\kappa$ by an order of magnitude leads to large changes in the rate of damping, i.e the amplitude of longitudinal (acoustic) wave decreases rapidly with an increasing value of $\kappa$.

\begin{figure}[!h]
\centering
\includegraphics[height=0.5\textheight,width=0.6\textwidth]{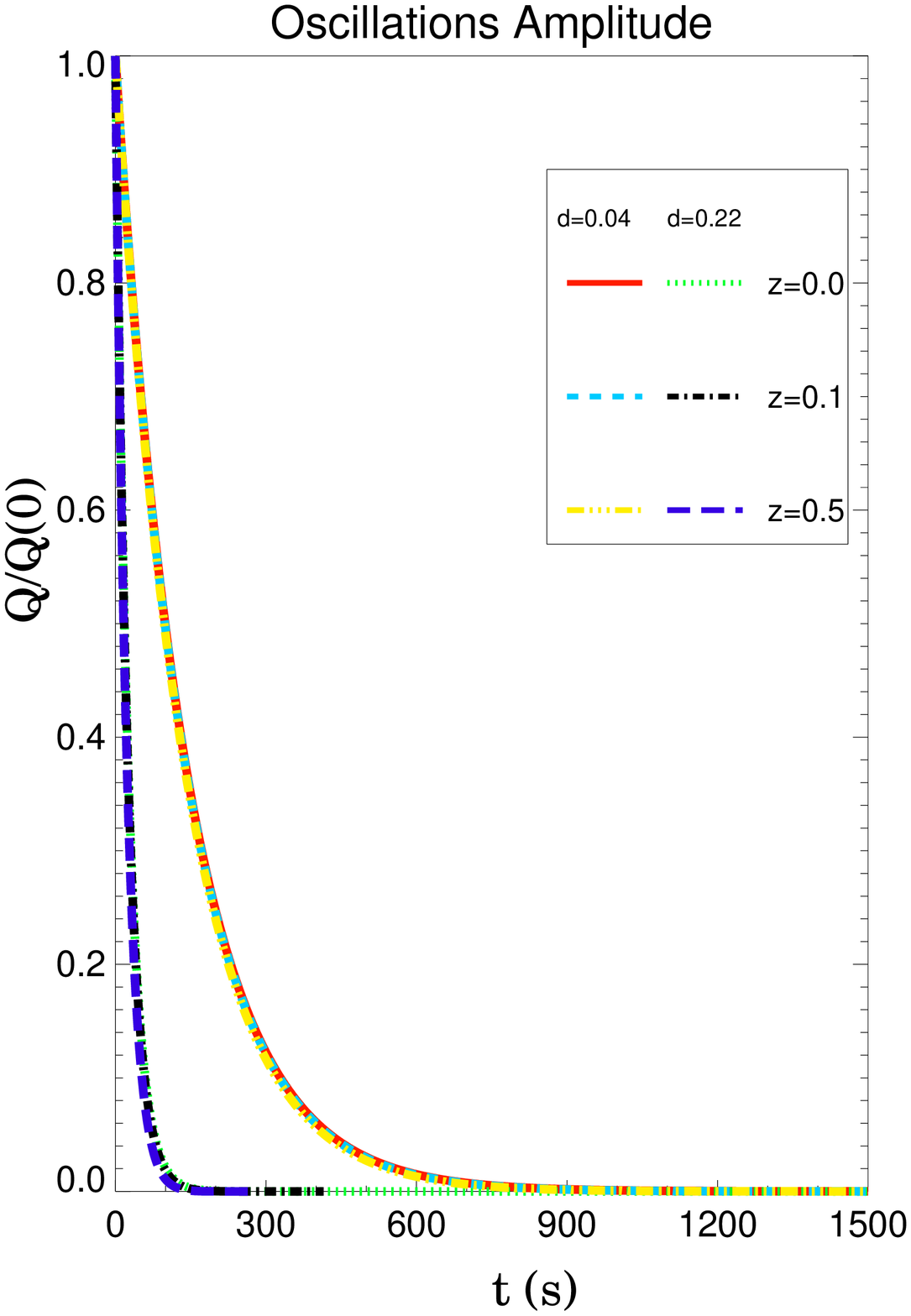}
\vspace{-1 cm}
\caption{The graph shows the amplitude of oscillations as function of time at variable positions along the coronal magnetic field lines, e.g. $z=(0.0, 0.1, 0.5)$ and $\lambda=0.1$, and with different values of $d=(0.04, 0.22)$.}\label{Amp-damping_z}
\end{figure}

Next, Figure \ref{Amp-damping_z} shows how the decay of the amplitude varies for different values of $z=(0.0, 0.1, 0.5)$. Two values of thermal ratio $d=(0.04, 0.22)$ are taken to illustrate the influence of thermal conduction on cool and hot corona, $T$=(1 MK, 3 MK), and the characteristic value of $\lambda=0.1$. This plot shows that the strength of the damping due to thermal conduction will be for all intents and purposes, the same along a loop.

\section{Discussion and Conclusion}
The influence of variable background on the longitudinal (field-aligned) MHD wave propagating in a homogeneous, magnetised plasma in a weakly stratified atmosphere has been investigated. The background plasma is assumed to be cooling as the magneto-acoustic wave propagates. Thermal conduction is the dominant mechanism causing the plasma cooling and changing pressure as a function of time. The magnetic field was assumed to be constant and directed along the $z$ direction. A governing equation is derived under the assumption that the background plasma is cooling on a time scale greater than or comparable to the characteristic period of the perturbations. A time- and space-dependent dispersion relation that describes the propagation of the longitudinal magneto-acoustic mode is obtained and an analytic equation for the time-dependent amplitude is also derived. The effect of cooling background due to thermal conduction on the amplitude of hot loop oscillations was then studied.

The governing equation in terms of time and space is solved by using the WKB theory. Leading and first order equations for the dispersion relation and wave amplitude, respectively, are obtained and solved analytically. An approximate solution which represents the properties of the field-aligned acoustic wave is found with the aid of method of characteristics. Numerical evaluations illustrate the behaviour of the analytic solutions. A comparison of wave behaviour for a range of initial temperatures is studied.

Enhanced by the cooling background plasma, the amplitude of longitudinal (acoustic) waves was found to rapidly decay due to the influence of thermal conduction. The rate of damping of the oscillations was found to depend on the initial temperature of the plasma and the amount of stratification. It was previously shown that thermal conduction leads to the damping of slow mode oscillations.  For example, the amplitude of slow modes in \inlinecite{Moortel03} was detected to undergo damping whereas in this study the amplitude was found to experience a much stronger damping. It should be noted that \inlinecite{Morton09d} reported that the damping of the slow mode was also heavy where the amplitude of wave suffers a rapid decrease due to the cooling of the background plasma by another dissipation, namely by radiation mechanism.  Consequently, we conclude that the magneto-acoustic oscillations of hot corona can be exposed to a strong damping because of cooling by thermal conduction while radiation is a dominant method of damping cool coronal oscillations.

Overall, the presented results into the study of the damping of coronal oscillation imply that both radiation and thermal conduction mechanisms should be taken into account even for background plasma as a result of their efficient effect on dominating the properties and lifetimes of MHD oscillations in cool and hot corona, respectively. The main point is that these dissipative processes introduce a dynamic plasma background. Such background in often reported in observations. The MHD wave theory of dynamic plasma is in its infant stage and we argue that for an adequate modelling of the solar coronal processes such studies as consideration of dynamic plasma background in the modelling is essential. The temporal evolution of the background plasma has an extremely important effect on the properties of waves and needs to be incorporated into future models. This is especially necessary if deriving plasma parameters from magneto-seismological methods, e.g. in applications to the solar corona.

\medskip
\begin{Ack}
The authors would like to thank Prof M.S. Ruderman for useful discussion. R.E. acknowledges M. K\'{e}ray for patient encouragement. The authors are also grateful to NSF, Hungary (OTKA, Ref. No. K67746), Science and Technology Facilities Council (STFC), UK and Ministry of Higher Education, Oman for the financial support.
\end{Ack}
\bibliographystyle{spr-mp-sola}

\end{article}
\end{document}